\documentstyle[psfig]{kapproc} 

\let\footnote\savefootnote

\let\footnotetext\savefootnotetext 

\let\footnoterule\savefootnoterule 

\kluwerbib

\startauthorindex

\newcommand{\dt}{$\Delta(\theta)$\ }
\newcommand{\ddt}{$\Delta(\theta)$}

\begin{document}

\articletitle{Polarization of Herbig Ae/Be\\
candidates and their environment}

\author{Cl\'audia V. Rodrigues}
\affil{Instituto Nacional de Pesquisas Espaciais - Brazil}
\email{claudia@das.inpe.br}

\author{Mar\'\i lia J. Sartori}
\affil{Laborat\'orio Nacional de Astrof\i sica - Brazil}

\author{Jane Gregorio-Hetem, A. M\'ario Magalh\~aes}
\affil{Instituto de Astronomia, Geof\'\i sica e Ci\^encias Atmosf\'ericas/USP -
Brazil}

\author{Celso Batalha}
\affil{Observat\'orio Nacional - Brazil}

\chaptitlerunninghead{Polarization of HAeBe stars}

\anxx{Author1\, and Author2}

\begin{abstract}
We present the V band polarization of 81 Herbig Ae/Be (HAeBe) candidates from
the Pico dos Dias Survey (PDS). A good estimate of the foreground polarization
was possible for most stars. A large number of objects shows intrinsic
polarization, which indicates that their circumstellar envelopes have some kind
of non-spherical symmetry. In this work, however, we study these data focusing
on their relation with the interstellar medium. Our data seem to indicate a
correlation between the position angle of the HAeBe star polarization and that
of the corresponding field stars. This may be an evidence that the ambient
interstellar magnetic field can play a role in the development of asymmetries
in the envelope of young stars of intermediate mass. The spatial distribution
of the sample relative to neighboring star forming regions is also studied.

\end{abstract}

\section*{Introduction}

Herbig Ae/Be (HAeBe) are pre-main-sequence stars of intermediate mass.
Differences in physical properties, evolution, and circumstellar material
allow a separation in two
sub-groups: Ae, spectral type later than or equal to B6;
Be, spectral type earlier than or equal to B5 (Natta et al. 2000). In this
colloquium, it was also proposed another classification based on their 
spectral energy distribution (Sartori et al. 2003).
From Group 1 to 4, the fraction of the total flux emitted at infrared
wavelengths increases.

The Pico dos Dias Survey (PDS -- Gregorio-Hetem et al. 1992; Torres et al. 1995;
Torres 1999) has found 105 HAeBe candidates. We present the polarization of a
large portion of the above sample (81 objects) and study possible correlations
with the interstellar magnetic field. The study of the polarization in the
context of the circumstellar material properties will be done elsewhere. We
also present the spatial distribution of the HAeBe candidates relative to dark
clouds.

\section{Polarization Data}

The observations have been done with the 0.60-m Boller \& Chivens telescope at
the Observat\'orio do Pico dos Dias, Brazil, operated by the Laborat\'orio
Nacional
de Astrof\'\i sica, Brazil, from 1998 to 2002. It was used a CCD camera modified
by
the polarimetric module described in Magalh\~aes et al. (1996).
The reduction has been done following the standard steps of differential
photometry using the IRAF facility\footnote{IRAF is distributed by National
Optical Astronomy Observatories, which is operated by the Association of
Universities for Research in Astronomy, Inc., under contract with the National
Science Foundation.}. The polarimetric reduction was greatly facilitated by the
used of the package {\it pccdpack} (Pereyra 2000).

\section{Foreground polarization}

The observed polarization is, in the general case, composed by two components:
the intrinsic (to the object) polarization plus a foreground polarization
(produced in the intervening interstellar medium). So the first step to study
the polarization of a object is to estimate its foreground polarization which
must be subtracted from the observed polarization.
The foreground polarization is estimated by the weighted average of the
polarization of the field stars in the CCD images. The number of objects in each
field vary from
4 to more than 1000: the average number is 175.

\section{Envelope asymmetry $\times$ interstellar magnetic field}

Considering the standard mechanism of grain alignment in the interstellar
medium, the
foreground polarization traces the magnetic field direction. The intrinsic
polarization gives us a direction related to the asymmetry of the envelope.
We can ask if, in our sample, these two directions have some kind of
correlation.
For this, we define \dt as the angle difference between the intrinsic
polarization and the foreground polarization.

\begin{figure}[ht]
\psfig{file=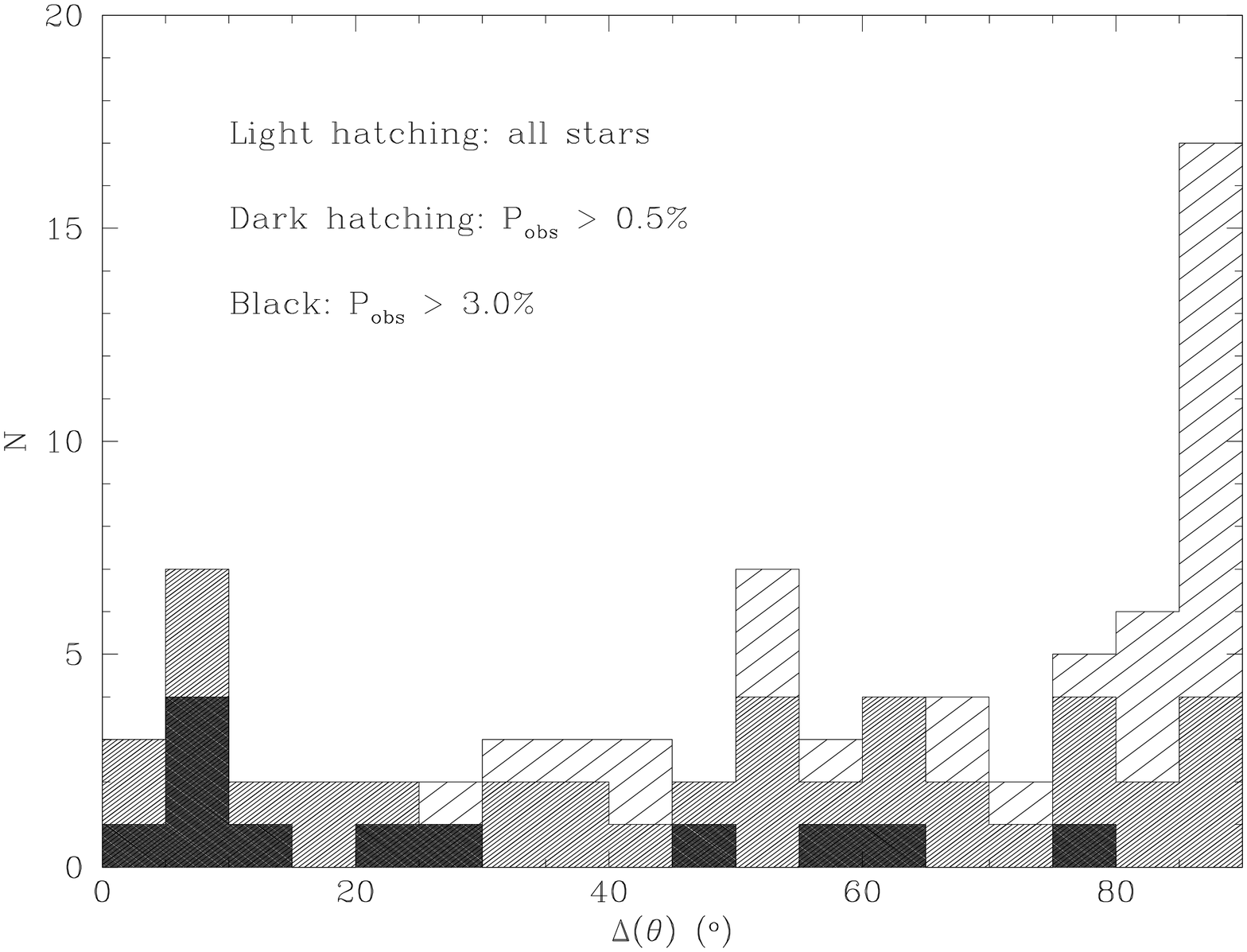,height=2.40in,angle=-90}
\vskip -2.4in
\narrowcaption{Histogram of the \dt for the HAeBe sample.}
\vskip 1.6in
\end{figure}

Figure 1 shows the histogram of \ddt. When all objects are considered (light
hatching) , there is
clear peak around 90$^{\rm o}$. However, an inspection of the data shows that
this peak is associated with objects with small values of observed
polarization. In this case, the intrinsic polarization will have the same
modulus of the foreground polarization, but perpendicular to it: causing the
peak observed at 90$^{\rm o}$. Probably these candidates are at small distances
and the polarization of field stars - at larger distances than the candidate -
represents a background (not a foreground) polarization. In that case, the
foreground polarization must be negligible for these HAeBe candidates stars.

For the above reason, we also present the histogram considering only objects
having the observed polarization larger than 0.5\% (dark hatching). These
objects may be less affected by the foreground correction, since they must have
a larger contribution from the intrinsic polarization to the observed value.
The peak at 90$^{\rm o}$ disappears. A smaller peak at 0$^{\rm o}$ remains:
this peak is also present for the stars with observed polarization larger than
3.0\% (black) as well as for intrinsic polarization larger than 3.0\%
(not shown in Figure 1). This excess of objects with \dt around zero - 5\%
significant - could mean that the ambient interstellar magnetic field can play
a role in shaping the envelope. The peak at 0$^{\rm o}$ does not exist if we
separate the stars in spectral types or in SED groups, probable because the
small number of objects.

\section{Spatial distribution of the sample}

Figure 2 shows the positions of the 105 HAeBe candidates in Galactic coordinates
and the positions of the dark clouds from Lynds (1962) and Feitzinger \& Stuwe
(1984) nearest to the HaeBe candidates. 

\begin{figure}[ht]
\psfig{file=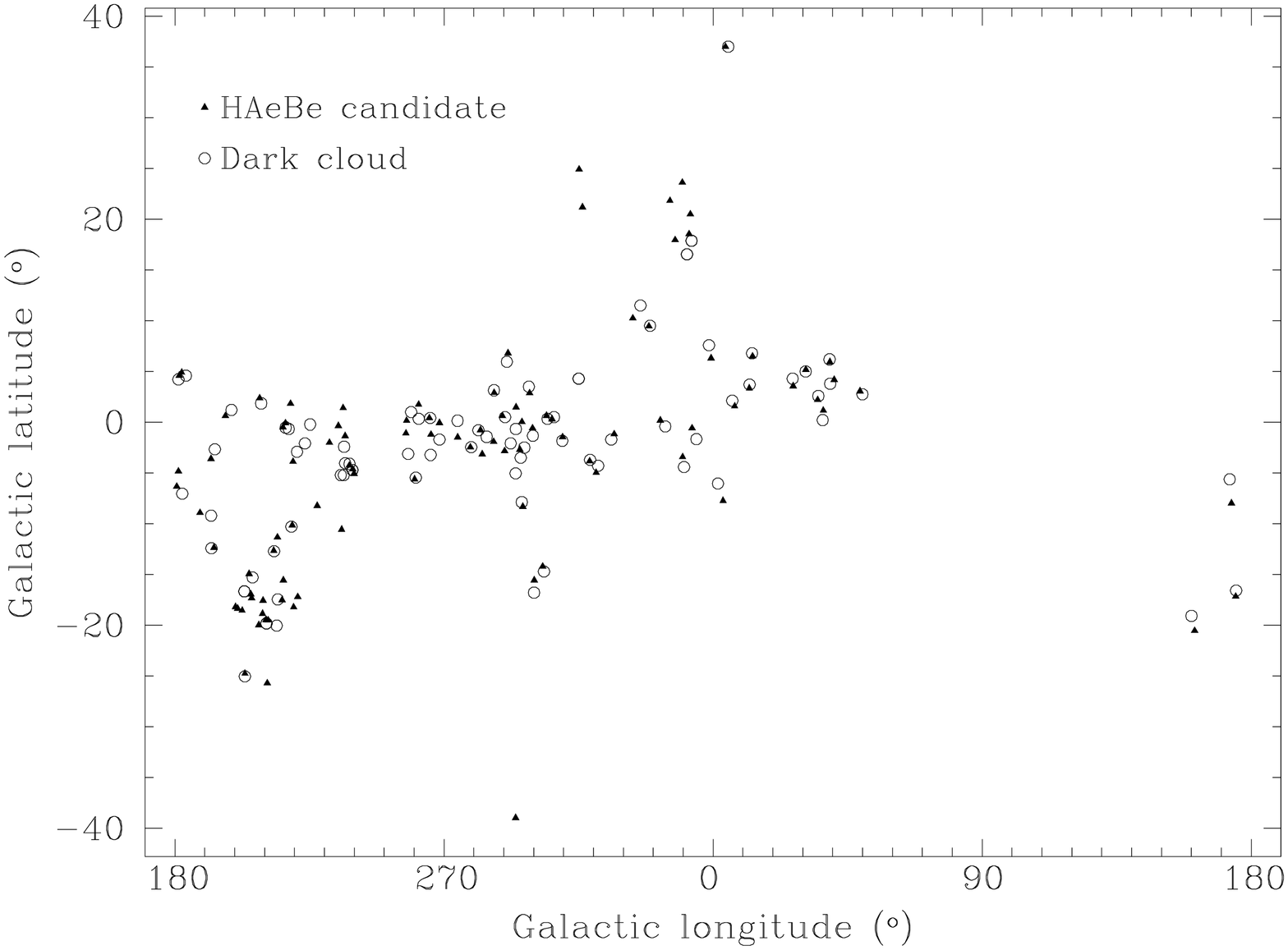,height=2.4in,angle=-90}
\vskip -2.4in
\narrowcaption{Spatial distribution of HAeBe candidates and dark clouds.}
\vskip 1.8in
\end{figure}

\begin{figure}[ht]
\psfig{file=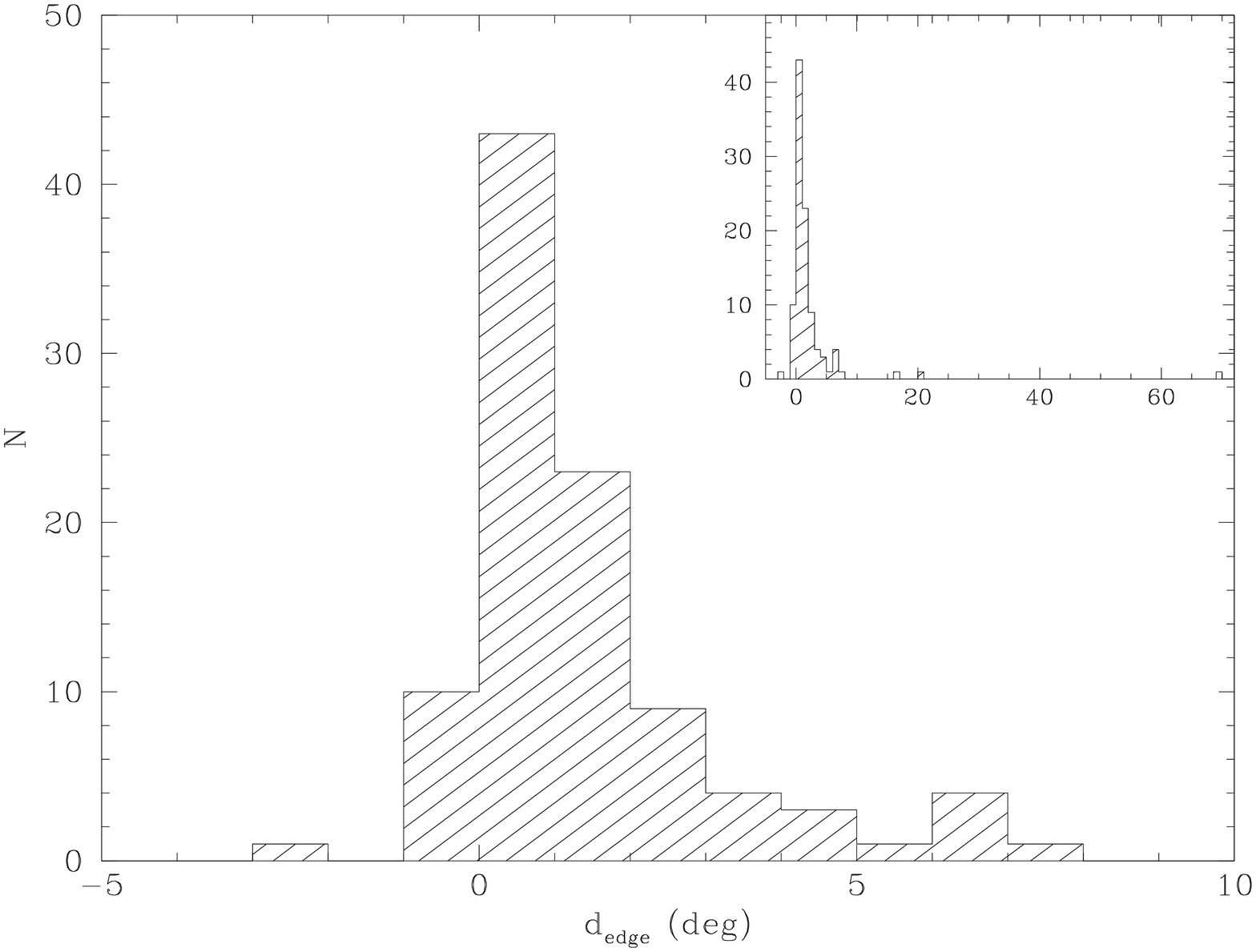,height=2.4in,angle=-90}
\vskip -2.4in
\narrowcaption{Frequency distribution of angular distances of the HAeBe
candidates to the edge of the nearest dark cloud.}
\vskip 1.8in
\end{figure}

The angular distance to the edge of the nearest cloud, $d_{edge}$, was
calculated for all HAeBe candidates. Negatives values of $d_{edge}$ are
obtained for objects seen in the cloud direction. A crude estimate to the error
is 1$^{\rm o}$. The distance histogram is shown in Figure 3. There is a clear
peak at $d_{edge}$ around 0.5$^{\rm o}$. A number of 54 (out of 105) objects
has $d_{edge}$ smaller than 1$^{\rm o}$.  Group 4 (IR dominated SED) has a 
$d_{edge}$ distribution statistically inconsistent with the other groups: the
sources in that group have apparently smaller $d_{edge}$ (Figure 4). This result
is consistent with the assumption that Group 4 contains the younger objects
(more embedded). No
significant statistical differences between the Ae and Be stars was found.

\vskip -0.2in
\begin{figure}[bht]
\psfig{file=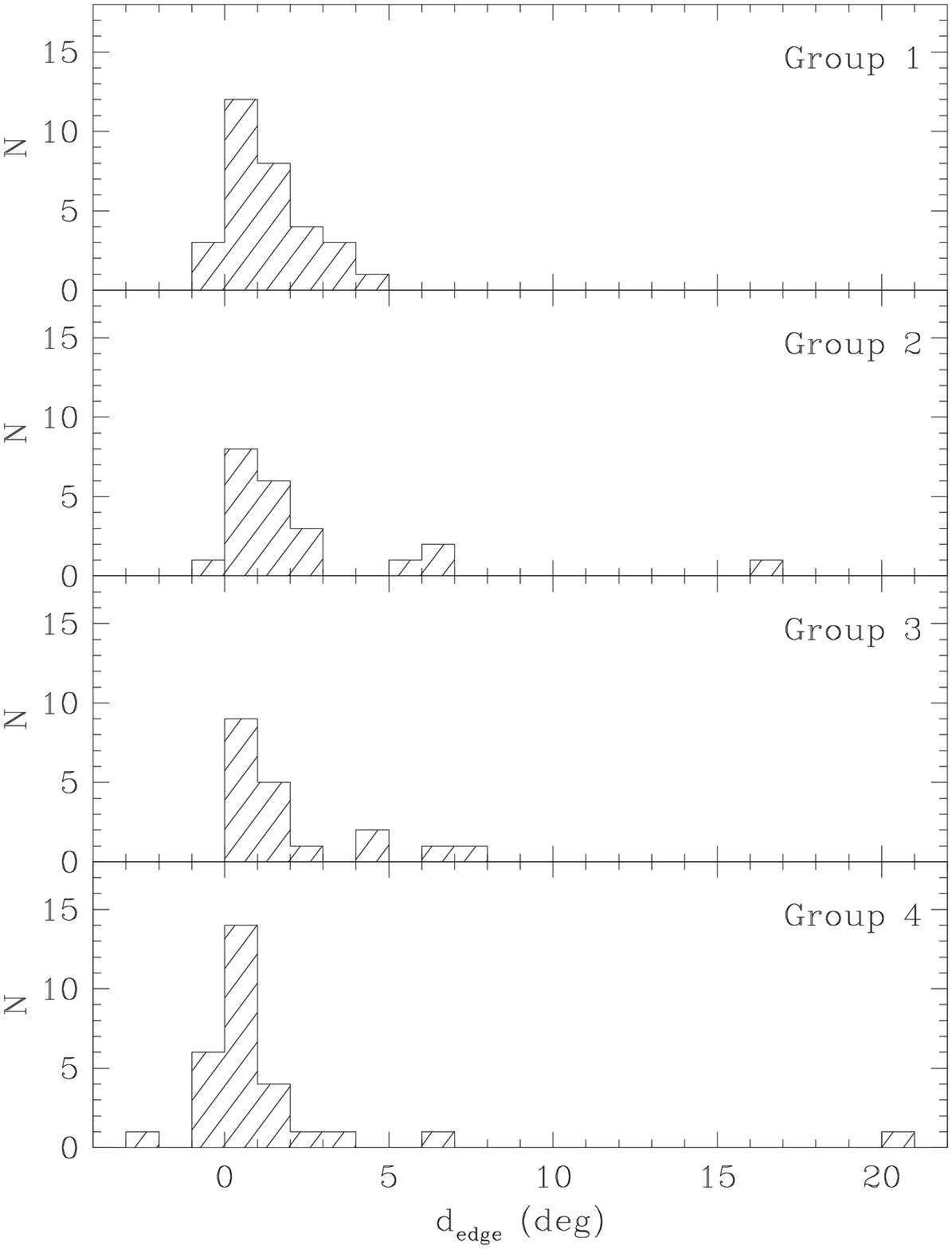,height=3.7in}
\vskip -3.7in
\narrowcaption{Frequency distribution of angular distances of the HAeBe
candidates to the edge of the nearest dark cloud grouped by SED class.}
\vskip 2.3in
\end{figure}

\begin{acknowledgments}
This work was partially supported by Fapesp (Procs. 2001/09018-2
and 2001/12589-1).
\end{acknowledgments}

\end{document}